\documentclass[prb,twocolumn,showpacs,preprintnumbers,amsmath,amssymb]{revtex4}

\usepackage{amsmath}
\usepackage{graphicx}
\usepackage{dcolumn}
\usepackage{bm}
\usepackage{natbib}
\usepackage[colorlinks=true,citecolor=blue,linkcolor=blue,urlcolor=blue]{hyperref}

\newcommand{\tb}{\ensuremath{\mathrm{TbFe_3(BO_3)_4}}}
\newcommand{\re}{\ensuremath{\mathrm{RFe_3(BO_3)_4}}}
\newcommand{\gd}{\ensuremath{\mathrm{GdFe_3(BO_3)_4}}}
\newcommand{\nd}{\ensuremath{\mathrm{NdFe_3(BO_3)_4}}}
\newcommand{\yfe}{\ensuremath{\mathrm{YFe_3(BO_3)_4}}}

\begin{document}

\preprint{CAPS/123-QED}

\title{Non-Resonant X-ray Magnetic Scattering on Rare-Earth Iron Borates RFe$_3$(BO$_3$)$_4$ }

\author{J.E. Hamann-Borrero$^1$}
  \email{j.e.hamann.borrero@ifw-dresden.de}	
\author{M. Philipp$^1$}
\author{O. Kataeva$^{1,2}$}
\author{M. v. Zimmermann$^3$}
\author{J. Geck$^1$}
\author{R. Klingeler$^{1,6}$}
\author{A. Vasiliev$^4$}
\author{L. Bezmaternykh$^5$}
\author{B. B\"{u}chner$^1$}
\author{C. Hess$^1$}
\affiliation{$^1$Leibniz Institute for Solid State and Materials Research, IFW Dresden, 01171 Dresden, Germany}
\affiliation{$^2$A.E.Arbuzov Institute of Organic and Physical Chemistry of the Russian Academy of Sciences, Arbuzov Str. 8, Kazan, 420088, Russia.}
\affiliation{$^3$Hamburger Synchrotronstrahlungslabor HASYLAB at Deutsches Elektronen-Synchrotron DESY, Notkestr. 85, 22603 Hamburg, Germany.}
\affiliation{$^4$Low Temperature Physics department, Faculty of Physics, Moscow State University, Moscow, 119992 Russia.}
\affiliation{$^5$L.V Kirensky Institute of Physics, Siberian Division, Russian Academy of Sciences, Krasnoyarsk, 660,0,36 Russia.}
\affiliation{$^6$Kirchhoff Institute for Physics, University of Heidelberg, Im Neuenheimer Feld 227, D-69120 Heidelberg }
\date{\today}

\begin{abstract}
Hard x-ray scattering (HXS) experiments with a photon energy of 100keV were performed as a function of temperature and applied magnetic field on selected compounds of the \re~ family. The results show the presence of several new diffraction features, in particular non-resonant \textit{magnetic} reflections in the magnetically ordered phase, and structural reflections that violate the diffraction conditions for the low temperature phase $P3_121$ of the rare-earth iron borates. The temperature and field dependence of the magnetic superlattice reflections corroborate the magnetic structures of the borate compounds obtained by neutron diffraction. The detailed analysis of the intensity and scattering cross section of the magnetic reflection reveals details of the magnetic structure of these materials such as the spin domain structure of \nd~ and \gd. Furthermore we find that the correlation length of the magnetic domains is around 100 \AA{} for all the compounds and that the Fe moments are rotated $53^\circ\pm3^\circ$ off from the hexagonal basal plane in \gd.
\end{abstract}
\pacs{Valid PACS appear here}

\maketitle

\section {Introduction}
The family of rare-earth compounds with the chemical formula \re~(R = Rare Earth or Y) has triggered  a considerable attention in the last few years. From a fundamental point of view, already the presence of two different magnetic ions ($3d$ and $4f$ elements) which form two interacting magnetic sub-lattices, suggests a subtle interplay of complex magnetic ground states. In addition, a rich variety of interesting structural and dielectric properties has been observed in these materials, partially coupled to the systems' magnetism, which is evidenced by a plethora of structural and magnetic phase transitions which depend on the rare-earth ion \cite{vasiliev06a,tristan07,Popova07a,kadomtseva05,krotov06,fausti06,Vasiliev06,yen06}.
Furthermore, magneto-electric coupling and multiferroic features, i.e., the coexistence of elastic, magnetic, and electric order parameters have been reported for the Nd and Gd based compounds \cite{zvezdin05,kadomtseva06,zvezdin06,yen06}.

At room temperature \re~ compounds crystallize in the space group $R32$ \cite{Joubert}. For "light" rare-earth ions from La to Sm, this structure is kept until low temperatures. "Heavier" rare-earth ions from Eu to Yb, and also Y cause a symmetry reduction to space group $P3_121$ upon lowering temperature which is manifested as a sharp peak at the transition temperature in specific heat measurements \cite{Popova07a,vasiliev06a,popova08}. The transition temperature $T_S$ depends basically on the size of the R-type ion present in the structure\cite{fausti06} and one observes a decreasing $T_S$ by increasing the R radius. In particular, one finds $T_S=201.5$~K, 155~K, 445~K for the Tb, Gd, Y based compounds, respectively.\cite{fausti06} Note, that \nd~does not undergo the symmetry reduction and remains in the $R32$ space group.
The main elements of the crystal structure of the high symmetry $R32$ phase are spiral chains of edge-sharing $\rm FeO_6$ octahedra running along the $c$-axis. Each rare-earth ion is coordinated by six oxygen ions forming a triangular $\rm RO_6$ prism. These prisms are separated from each other by regular $\rm BO_3$ triangles with no common oxygen ions. Both the $\rm BO_3$ triangles and $\rm RO_6$ prisms connect three $\rm FeO_6$ chains \cite{Joubert}. For the low symmetry phase $P3_121$ there are two nonequivalent iron positions, and one of the iron chains is shifted along the $c$-axis with respect to the other two chains in the unit cell\cite{klimin05}.\\

\begin{figure*}
\includegraphics[angle=-90,width=1.9\columnwidth]{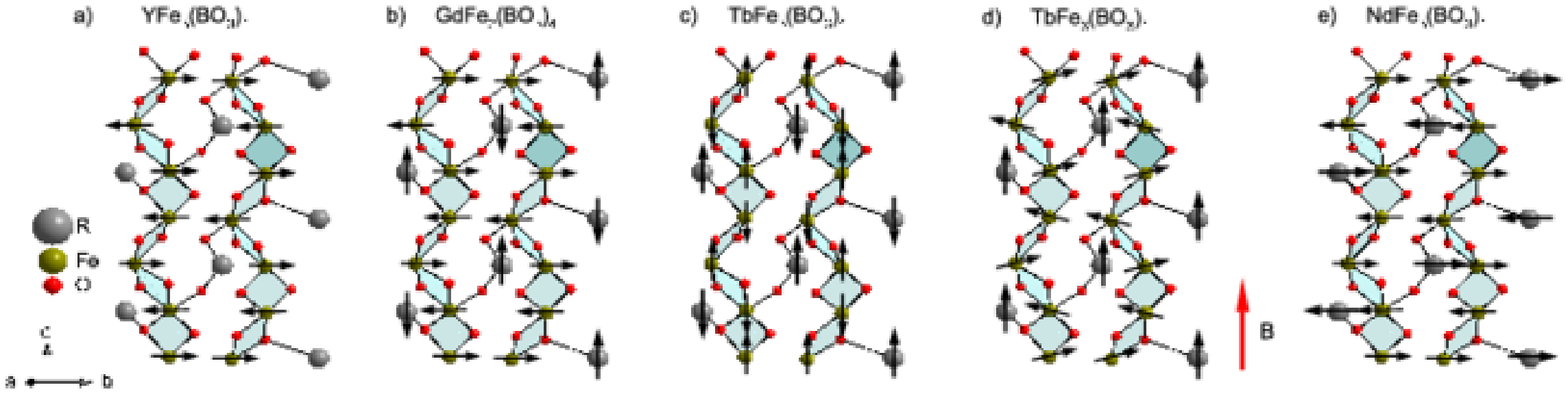}
\caption{Crystal structure of \re~showing the the magnetic structure for a) \yfe~below $T_N$, b) \gd~ in the temperature range $T_N>T>T_{\mathrm{SR}}$. Figure c) shows the spin structure for \tb~below $T_N$ and $B<3.5~T$. And \gd~ at $T<T_{\mathrm{SR}}$. Figure d) shows the magnetic structure of  \tb~ in the spin-flop state at applied magnetic fields larger than 3.5~T at $T=2$~K. Figure e) shows the spins structure of \nd~.  In the plots only two iron chains are plotted and the boron ions were removed for clarity.}
\label{fig:spins}
\end{figure*}

Regarding magnetic ordering, several interesting features are present in these materials. Previous measurements of magnetization \cite{Popova07a}, specific heat\cite{vasiliev06a} and other techniques like M\"{o}ssbauer \cite{hinatsu03} spectroscopy and infrared absorption spectroscopy \cite{chukalina04} have revealed a second order antiferromagnetic (AFM) ordering transition of the iron sub-lattice at low temperatures (in the range  $\sim 30$~K to 40~K). The orientation of the Fe moments depends on the rare-earth ion present in the structure. More specifically, at low temperature ($\sim 2$~K) the Fe moments lie within the $ab$ plane\cite{ritter08,fischer06,popova-2010} for the Y and Nd based  compounds,  while for the Gd and Tb based they are parallel to the $c$-direction (figure \ref{fig:spins}c). Furthermore, a first order magnetic phase transition is present in the Tb, Dy and Gd compounds, where the antiferromagnetically ordered iron moments undergo a spin flop from the easy axis state along the $c$-direction to an easy plane one along the $ab$ plane when a magnetic field along the $c$-axis is applied at low temperature.\cite{balaev03,Popova07a} In the case of \tb~ the Fe spin flop is accompanied by a reconfiguration of the Tb moments from an antiparallel to a parallel arrangement (cf. Fig. \ref{fig:spins}d)\cite{Popova07a,ritter07}. Interestingly, in the case of \gd~ there is also a temperature-driven reorientation of the Fe moments from $c$-axis orientation to an $ab$-plane one, which occurs already in zero magnetic field at $T_\mathrm{SR}=9.3$~K. Here, the Gd moments are polarized along the $c$ axis by a biasing internal magnetic field created as a result of the Fe$^{3+}$-Gd$^{3+}$ exchange interaction.\cite{pankrats,yen06} The resulting spin structure is sketched in figure \ref{fig:spins}b.

Recently \gd~ and \nd~ have been reported to exhibit a significant magneto-dielectric coupling\cite{zvezdin05,kadomtseva06}. For electric polarization to appear, small distortions or displacements of the atoms from their symmetry position are necessary. Therefore for the magnetically induced polarization present in these samples, a lattice distortion is expected which has not yet been seen experimentally. Magnetostriction measurements\cite{krotov06,kadomtseva06} have shown a clear relation between lattice and magnetoelectric properties, but a proper description of the lattice distortions that produce the observed polarization is still lacking. 
In recent neutron scattering investigations which confirm the afore described spin structures, no new information is obtained regarding structural displacements which would explain the observed electric polarization, mainly due to the low \textbf{q} resolution of neutrons as compared to x-rays. Moreover it is not clear if the observed superstructures are purely magnetic or also structural. In this respect detailed x-ray diffraction studies are necessary to elucidate small structural and magnetic features.

In this paper we present a comprehensive hard x-ray scattering study on \re~ with R$ = $Gd, Tb, Nd and Y in order to better understand the structural and magnetic properties of the material. Our data reveal a weak superstructure reflection at $(0,0,3l\pm1.5)$, i.e., at the antiferromagnetic ordering vector seen in neutron scattering. This reflection is clearly correlated to an in-plane ordering of the Fe spins and -- depending on the material -- can thus be induced by an external magnetic field. Through a careful analysis of the $\bf q$ dependence of this reflection we rule out that it is related to a distortion of the lattice and demonstrate its purely magnetic character. Detailed analysis of the integrated intensities of the Bragg and superlattice reflections, together with their scattering cross section, allows us to determine the size of the spin component of the magnetic moments which are perpendicular to the scattering plane and their relative orientation with respect to the hexagonal basal plane.
Moreover, in the earlier reported $P3_121$ low-symmetry structural phase we observe additional reflections at $(0,0,3l\pm1)$ that violate the diffraction conditions of $P3_121$ and thus are suggestive of an overlooked symmetry reduction when indexing the crystal structure, or the appearance of a structural distortion that induces a lattice modulation.

The paper is organized as follows: In section \ref{sec:magn-diff} a short introduction into x-ray magnetic scattering (XMS) is given followed by a discussion of the non-resonant scattering cross section of x-rays by magnetic materials at high photon energies, in the frame of our experimental setup. Experimental details are given in section \ref{sec:exp}, while the results of the observations and their discussion are presented in section \ref{sec:results}. This section is divided in two parts. One concerns the structural phase transition while the second part focuses on the magnetic ordering of the system, which causes a superlattice peak with Miller indices (0,0,1.5). The nature of which is demonstrated to be magnetic. Finally the work is summarized in section \ref{sec:summary}.

\section{X-ray magnetic diffraction}
\label{sec:magn-diff}

Neutron diffraction has been, since the determination of the magnetic structure of MnO by Shull and coworkers\cite{shull1951}, the primary tool for revealing the magnetic structure of magnetic materials. The main reason for this is the direct interaction between the neutron dipolar moment with the atomic magnetic moment of the sample. The cross section of this is comparable with the neutron-nuclei interaction, and thus explains comparable intensities of magnetic and nuclear reflections. On the other hand the intensity of an x-ray  beam scattered by unpaired electrons in a sample is much smaller than pure charge scattering by a significant factor\cite{blume85} as is demonstrated in the following equation:
\begin{equation}
\dfrac{\sigma_{mag}}{\sigma_{charge}}\simeq\left(\dfrac{\hbar\omega}{mc^2}\right)^2\dfrac{N^{2}_{m}}{N^2}\langle s\rangle^2\dfrac{f^{2}_{m}}{f^2}\sim 10^{-6},
\label{eq:magn-charg-ratio}
\end{equation}
with $N_m$ the number of magnetic electrons per atom, $N$ the number of electrons per atom. $f_m$ and $f$ are the magnetic and charge form factors, respectively. $\langle s\rangle$ is the expectation value of the spin quantum number and $\hbar\omega$ and $mc^2$ are the photon energy and electron's rest mass, respectively. Apart from using resonant x-ray magnetic scattering (RXMS), which has advanced to become a very successful tool in the past years\cite{Paolasini,nandi,mo}, one can overcome this difference in intensities only by using high brilliance, collimated and polarized sources like synchrotron radiation. This so-called non-resonant x-ray magnetic scattering (NRXMS) has only rarely been applied in the past years to study magnetic structures\cite{Bergevin1972,bruckel,strempfer,Chatterji2004}. However, compared to neutron scattering this technique offers some advantages since it is possible to separate spin and angular momenta from the scattered intensities. This intriguing property is associated with the matrix elements in the non-resonant cross section\cite{blume88}, which depend in different ways on the scattering geometry, photon energy and initial and final polarization states of the x-ray beam, as will be discussed in detail below.

\subsection{The Non-Resonant X-ray Magnetic Scattering Cross Section}
\label{sec:cross-sect}

\begin{figure}
 \includegraphics[angle=-90,width=0.7\columnwidth]{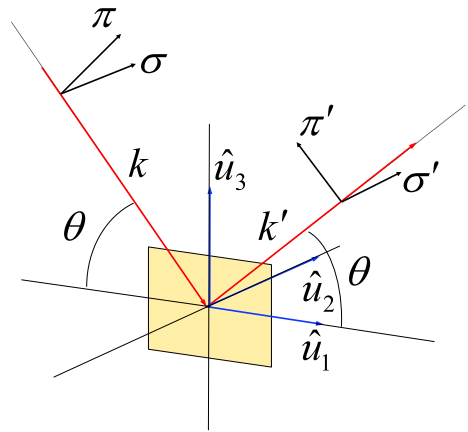}
 \caption{Definition of the coordinate system used to resolve the components of the spin moments of the sample as well as the polarization states of the incident and scattered photon.}
 \label{fig:cross-section}
\end{figure}

Out of various discussions of the non-resonant cross section of x-ray magnetic scattering\cite{blume88,bruckel,strempfer} we follow the one of Br\"{u}ckel et al.\cite{bruckel}. The elastic cross section for scattering of photons with initial incident polarization $\varepsilon$ and final polarization $\varepsilon^{'}$ can be written as
\begin{equation}
\left(\dfrac{d\sigma}{d\Omega}\right)_{\varepsilon\rightarrow\varepsilon^{'}}=r_{e}^{2}\vert \langle M_C\rangle_{\varepsilon^{'}\varepsilon}+i\dfrac{\lambda_c}{d}\langle M_M\rangle_{\varepsilon^{'}\varepsilon}\vert^2,
\label{eq:cross-section}
\end{equation}
where $r_{e}$ is the electron classical radius and $\lambda_c$ is the Compton length of the electron. The matrices $\langle M_M\rangle$ and $\langle M_C\rangle$ describe the polarization dependence of the magnetic and charge scattering amplitudes, respectively.
Figure \ref{fig:cross-section} depicts the reference frame for the diffraction experiments performed in this work. Here, $k$ and $k^{'}$ refer to the incoming and outgoing x-ray beams respectively. The $\sigma$ and $\pi$ vectors correspond to polarization perpendicular and parallel to the scattering plane. 
Finally, $\hat{u}_1$, $\hat{u}_2$ and $\hat{u}_3$ are the unitary vectors of the reference frame as defined in ref.~\onlinecite{bruckel}. Note that $\hat{u}_3$ is parallel to $k-k^{'}=q$, which defines the scattering vector. 
If we only consider pure magnetic diffraction ($\langle M_C\rangle=0$), in the frame sketched in figure \ref{fig:cross-section}, and taking into account that our diffraction measurements were performed with $\pi$-polarized incoming photons (see figure \ref{fig:cross-section}) at an energy of 100~keV ($\lambda=0.1239$~{\AA}). The matrix that describes the magnetic scattering amplitude $\langle M_M\rangle$ reduces to \cite{strempfer}:
 \begin{equation}
\langle M_M\rangle=\,
\begin{array}{c|cc}
 & \sigma & \pi \\ \hline
\sigma^{'} & S_2  & 0 \\
\pi^{'} & 0 & S_2 
\end{array}
\label{eq:magnetic-matrix-simple}
\end{equation}          
Where $S_2$ is the projection of the spin moment in the $\hat{u}_2$ direction.
This means that under high energy diffraction conditions, non-resonant magnetic reflections can be observed at positions of the reciprocal space where no charge scattering is present, if the spin moments of the atoms have a component perpendicular to the x-ray scattering plane.  Note that the incoming/outgoing beam polarization plays no role for the analysis of the diffracted intensities. Therefore, the cross section of a pure magnetic reflection under the above discussed conditions has the following form:
\begin{equation}
\left(\frac{d\sigma}{d\Omega}\right)_{m}=r_{e}^{2}\left(\frac{\lambda_{c}}{d}\right)^{2}|S_{2}|^{2}
\label{eq:mag-cross-section}
\end{equation}

\section{Experimental details}
\label{sec:exp}
Single crystals of \re~ with R = Y, Nd, Gd, Tb have been grown using a $\rm K_2Mo_3O_{10}$-based flux\cite{balaev03,bezmaternykh} and characterized by specific heat and magnetic susceptibility measurements using a physical property measurement system and a SQUID magnetometer, respectively, from Quantum Design\cite{tristan07,popova-2010,popova07,Popova07a}.
High energy x-ray diffraction experiments were performed  at beamline BW5 in Hamburg (HASYLAB at DESY) using an incident photon energy of 100~keV. The penetration depth of the x-rays at this energy is of the order of millimeters, enabling the study of bulk properties of large single crystals. The triple-axis diffractometer is equipped with a cryomagnet mounted on a double-tilt table with eulerian geometry and a solid state Ge detector (energy resolution of 500~eV at 100~keV)\cite{bouchard}. The sample was mounted inside the cryomagnet where temperatures down to 1.5~K can be reached and horizontal magnetic fields up to 10~T can be applied parallel and perpendicular to the scattering vector. The samples were aligned with the horizontal scattering plane being perpendicular to the $ab$ plane of the samples. Magnetic fields up to 8~T were applied along the $c$-direction and perpendicular to it.

The full integrated intensity of the observed reflections was extracted from the raw data by performing reciprocal lattice scans (e.g $l$-scans) at a given reflection, followed by scans along the $\omega$-direction at the maximum intensity of the former scan. The $l$ and $\omega$-scans were made extended to the peak sides until a constant background was reached. The scans were fitted using a Gaussian profile. 

\section{Results and discussion}
\label{sec:results}

\subsection{Structural Transition at T$_S$}

\begin{figure*}
 \includegraphics[angle=-90,width=1.5\columnwidth]{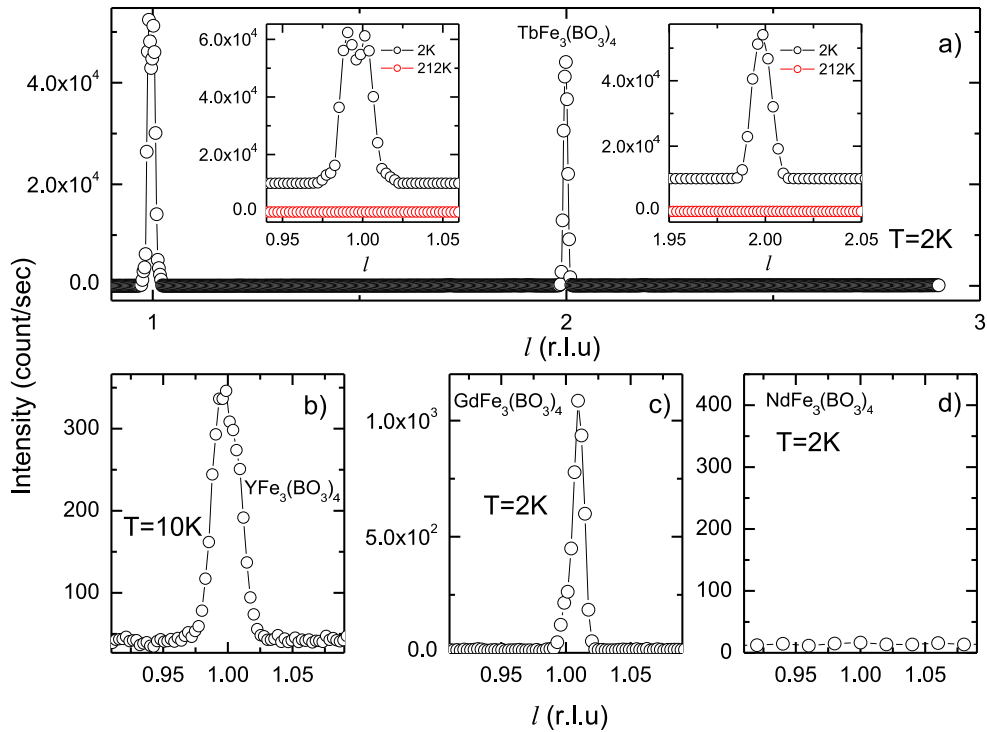}
 \caption{a) Scan at T=2 K $<<$ T$_S$ along the (0,0,$l$) direction for \tb~ showing the new superlattice reflections at $(0,0,3l\pm1)$. For clarity the (0,0,3) peak is not shown since it does not fit onto the scale. Insets in a) show that these reflections are absent in the high temperature phase $R32$. The high temperature data are shifted by 10$^4$ counts for clarity. b) and c) present $l$-scans showing the $(0,0,1)$ reflection at temperatures below $T_S$ for \yfe~ and \gd, respectively. d) In contrast \nd exhibits only the (0,0,3) reflection  at all temperatures.}
 \label{fig:lscans}
\end{figure*}

In order to search for superlattice reflections in our diffraction experiments, we performed overview scans along $(0,k,0)$ with  $0.45<k<2.9$ and along $(0,0,l)$ with $0.9<l<2.9$ at various temperatures. Figure~\ref{fig:lscans}a shows a representative example of the $(0,0,l)$-scan for \tb~ at low temperature (2~K) and at 212~K (Figure~\ref{fig:lscans}a insets), i.e., in the reported $R32$ and $P3_121$ symmetries of the crystal structure, respectively. Along this direction, the reflection conditions of both symmetries are identical, viz. ($0,0,3l$) with integer $l$. 
As can be seen in the figure, the reflection conditions for the high symmetry space group $R32$ are perfectly fulfilled. Interestingly, this is not the case for the low temperature data, where clear superlattice peaks at ($0,0,1)$ and ($0,0,2)$ are observed, thus violating the reflection conditions of the $P3_121$ space group. 

\begin{figure}
 \includegraphics[width=\columnwidth]{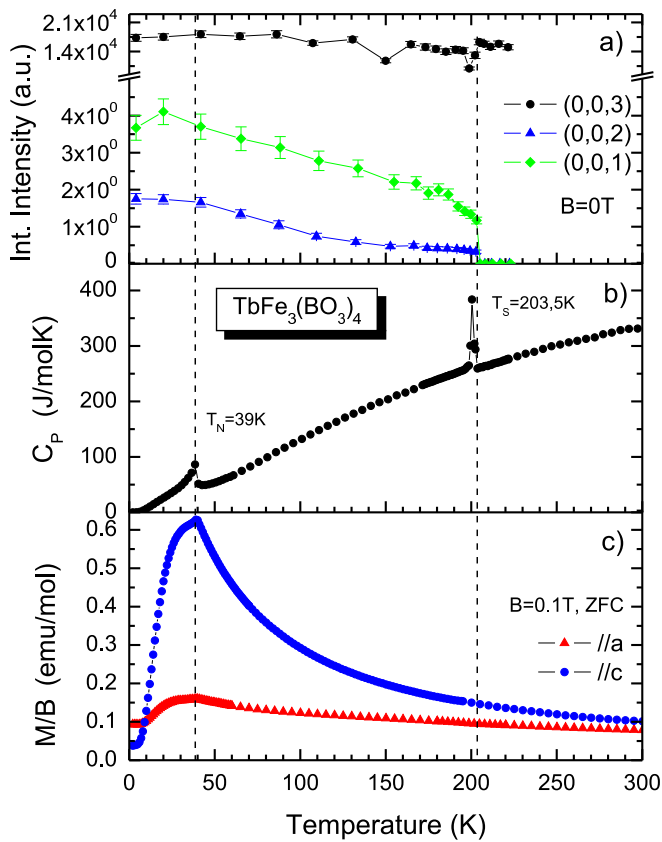}
 \caption{Structural and thermodynamic properties of \tb. a) shows the hard x-ray integrated intensities of the reflection ($0,0,1$), ($0,0,2$) and ($0,0,3$), b) Heat capacity. Plot c) shows the zero field cooled (ZFC) magnetic susceptibility when the field is applied parallel to $a$ and $c$, respectively, vs. temperature. Vertical lines denote the structural and magnetic phase transitions.}
 \label{fig:tb-all}
\end{figure}

We have studied the temperature dependence of these two unexpected reflections in comparison with the $(0,0,3)$ Bragg peak in detail as is shown in Fig.~\ref{fig:tb-all}. The data shown in part a) of the figure clearly reveal that the new superlattice peaks appear abruptly at the structural transition temperature $T_S=201.5$~K which is accompanied by a pronounced peak in the specific heat $c_p$ (cf. Fig.~\ref{fig:tb-all}b). Upon lowering the temperature, both reflections persist down to the lowest investigated temperature, where both peak intensities gradually increase without significant further changes, suggesting a stabilization of the structural distortion. The compound undergoes antiferromagnetic order at $T_N=39$~K as is evidenced\cite{Popova07a} by a further anomaly in the specific heat and a rapid decrease of the magnetic susceptibility (cf. Fig.~\ref{fig:tb-all}c). However, no significant change of the peak intensities occurs when crossing through this temperature.
The occurrence of these new, in the $P3_121$ symmetry forbidden reflections at $(0,0,3l\pm1)$ is not a particular feature of \tb, but it is a common feature of all \re~ compounds investigated in this study that experience a structural transition towards a symmetry lower than $R32$. This can be clearly seen in the panels b, c, and d of Fig.~\ref{fig:lscans}, which depict $l$-scans around $(0,0,1)$ for the Y, Gd, and Nd pendants of \tb. Indeed, pronounced peaks centered around $(0,0,1$) are observed for both the Y and Gd based compounds but not for \nd.

Our observation clearly implies that materials undergo a symmetry reduction at $T_S$ which results in a low temperature phase that is inconsistent with the $P3_121$ space group assigned so far. The new reflections at $(0,0,3l\pm1)$ have not been reported earlier, which is probably due to the fact that their intensities are about four orders of magnitude weaker than that at $(0,0,3)$ and thus can easily be overlooked unless the experiment relies on single crystals and x-rays from high brilliance sources like synchrotron radiation which probe the crystal's bulk. 

No possible symmetry subgroup of the $R32$ space group yields conditions that satisfy the presence of the $(0,0,1)$ and $(0,0,2)$ reflections. There are only two possible reasons that could explain them. The first possibility is related to the spiral chains of octahedra which run along the $c$-axis and yield the three-fold screw axis ($3_1$) symmetry feature. The presence of the $(0,0,1)$ reflection can be interpreted as a superlattice reflection resulting from a slight deviation of the Fe atom positions from the symmetry positions generated by the screw axis, e.g., by a slight displacement in the vertical direction from the symmetry position, resulting in a modulation that requires a tripling of the unit cell in order to fully describe the symmetry of the structure. The second possible explanation is based on a special case of multiple diffraction. More precisely, the symmetry reduction to $P3_121$ could, in principle, allow multiple scattering by planes in other orientations within the crystal, with a condition that is not possible at the higher $R32$ symmetry. In general, multiple scattering requires special geometry conditions which are already violated at small changes of the azimuth angle.
In order to discard multiple scattering events, azimuth scans on the $(0,0,1)$ reflection were performed with the result that the reflection remains observable even at azimuth angles larger than 90$^\circ$. Since multiple scattering is rapidly suppressed by rotating the crystal along the azimuth direction, multiple scattering can be disregarded as the origin of  $(0,0,3l\pm1)$ reflections. Our data thus implies that the superlattice reflections come from small structural distortions, as previously discussed, and not from multiple scattering events. 
 
Coming back to \nd, this compound shows the largest magnetic induced electric polarization measured among the ferroborates. Since in this compound no structural transition is present, i.e. no $(0,0,3l\pm1)$ reflections are observed (cf. Fig. \ref{fig:lscans}d). Thus, we can easily rule out the distortions responsible for these reflections as the origin of the electric polarization present in some of these compounds.

\subsection{Superlattice reflections in the antiferromagnetic phase}

\begin{figure}
 \includegraphics[angle=-90,width=\columnwidth]{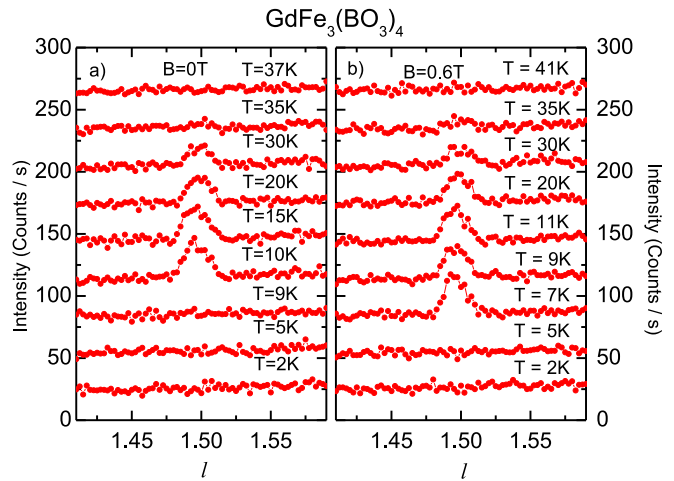}
 \caption{$l$-scans on \gd~ showing the magnetic (0,0,1.5) reflection at B=0 (a) and B=0.6T (b), as a function of temperature.}
 \label{fig:gd-lscans}
\end{figure}

\begin{figure}
 \includegraphics[width=\columnwidth]{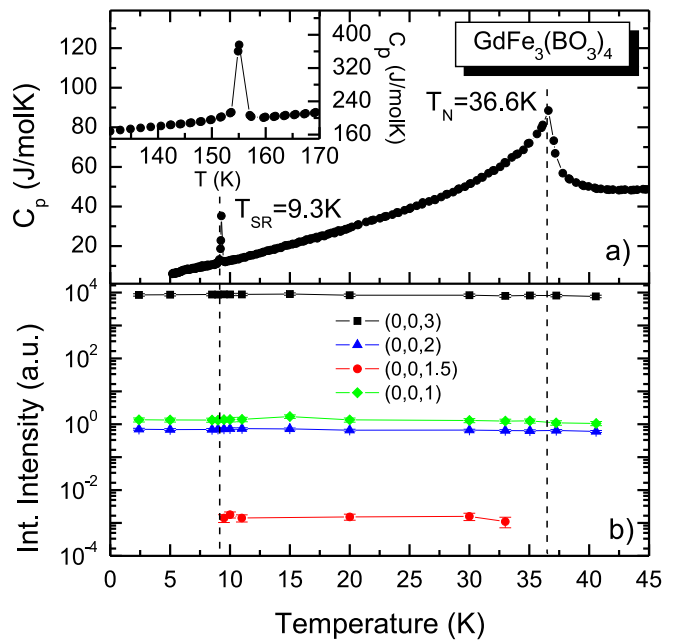}
 \caption{a) Heat capacity and b) hard x-ray integrated intensities measured on \gd. The super-lattice peak (0,0,1.5) is present only in the temperature region $T_{SR}<T<T_N$, where the iron spins are aligned in the $ab$-plane. The inset in figure a) shows the anomaly at the structural phase transition at $T_S=155$ K.}
 \label{fig:gd-all}
\end{figure}

Upon close inspection of the $(0,0,l)$ scans in the antiferromagnetically ordered phase ($T<T_N$) we observe further weak superlattice reflections at $(0,0,1.5)$ with an intensity about six to seven orders of magnitude smaller than the main $(0,0,3)$ reflection. A representative example for this reflection is shown in Fig.~\ref{fig:gd-lscans}a for the case of \gd.  Judging from the Miller indices of this reflection, the ordering of the magnetic ions generates an additional symmetry (super-lattice) which is commensurate with the chemical structure and involves a doubling of the unit cell along the $c$-direction. 
As can be seen in the data for \gd, the peak gradually emerges from the background a few degrees below $T_N=36.6$~K and persists below 30~K with constant intensity down to 10~K. Upon cooling further the peak abruptly vanishes and remains absent at $T\leq9$~K. Fig.~\ref{fig:gd-all} shows the temperature dependence of the peak intensity in comparison with the specific heat. Obviously, the disappearance of the peak occurs exactly at the spin reorientation transition at $T_\mathrm{SR}=9.3$~K where the Fe-spins turn from an easy-plane  configuration in $ab$ to an easy-axis one along the $c$-axis. This suggests that the superstructure which gives rise to the new superlattice reflections is related to the in-plane orientation of the iron spins.

\begin{figure}
 \includegraphics[width=\columnwidth]{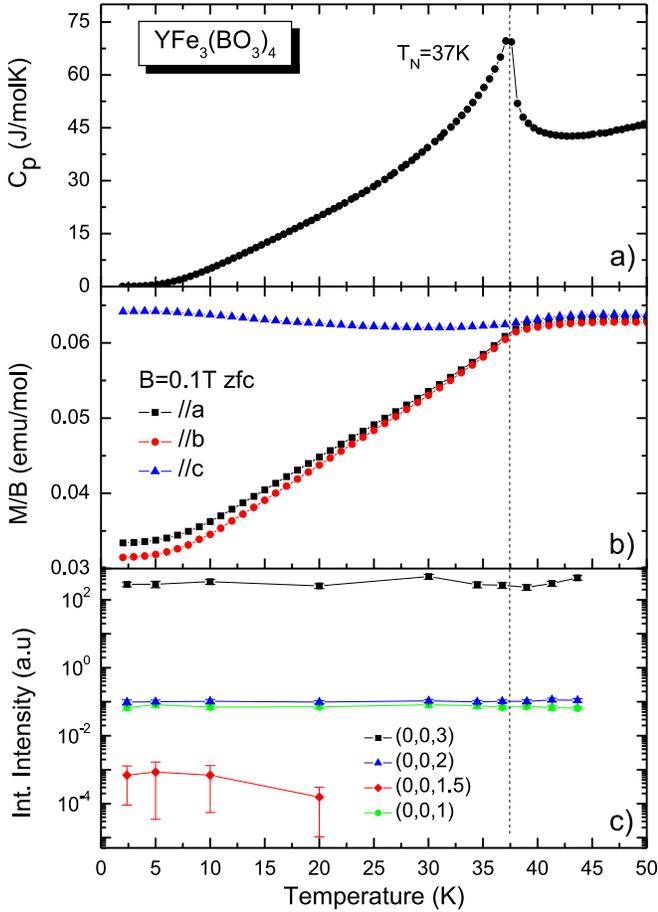}
 \caption{a) Heat capacity, b) ZFC static magnetic susceptibility and c) hard x-ray integrated intensities measured on \yfe. (0,0,1) and (0,0,2) reflections are present below $T_S$ and the super-lattice reflection (0,0,1.5) appears when the magnetic order of the samples evolves.}
 \label{fig:y-all}
\end{figure}

This conjecture is corroborated by further investigations on the Y, Nd and Tb based compounds which exhibit in-plane Fe spin order (Y, Nd) and  $c$-axis oriented moments (Tb), respectively. Fig.~\ref{fig:y-all} shows the temperature dependence of the specific heat, magnetic susceptibility and diffraction peak intensities for \yfe. The onset of antiferromagnetic order of the iron spin with $ab$-orientation below $T_N=37$~K can clearly be inferred from the strong peak in the specific heat at this temperature and the strong decrease of the in-plane magnetic susceptibilities\cite{hinatsu03}. Similarly as in \gd, the $(0,0,1.5)$ peak emerges below $T_N$. We point out, however, that the peak could be resolved only below 20~K.

\begin{figure}
 \includegraphics[width=\columnwidth]{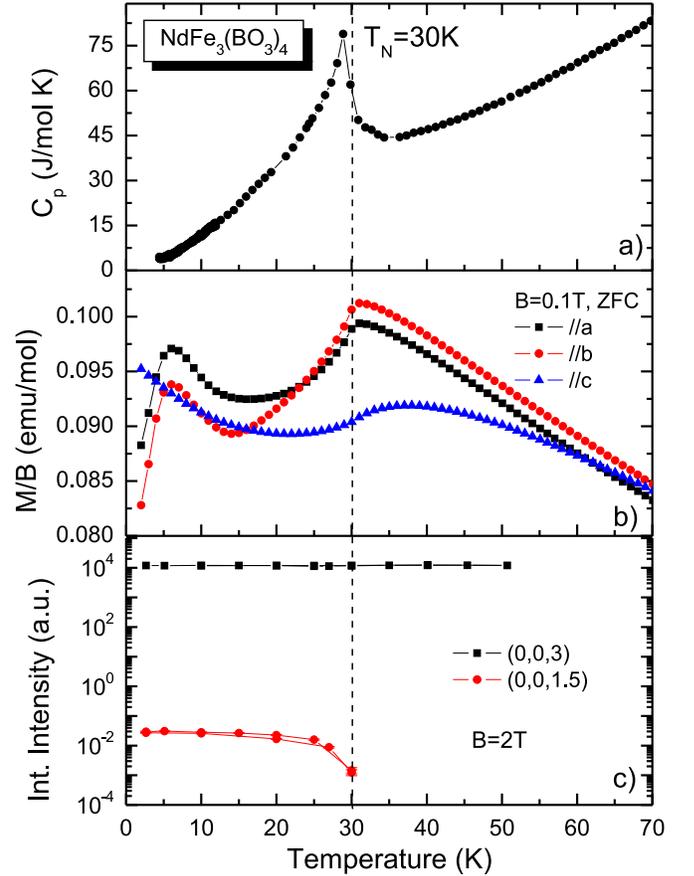}
 \caption{Low temperature behavior of \nd.~ a) Heat capacity, b) ZFC magnetic susceptibility and c) hard x-ray integrated intensities. The dashed line indicates the ordering temperature and shows the correlation between magnetic ordering of the sample and the appearance of the super-lattice peak (0,0,1.5).}
 \label{fig:nd-all}
\end{figure}

We find a very similar result for \nd~ as is shown in Fig.~\ref{fig:nd-all}. The magnetic ordering temperature $T_N\approx30$~K is somewhat reduced in this material as compared to the previous discussed cases. As in \yfe, the spin configuration is in-plane\cite{fischer06} as is signaled by the magnetic susceptibility shown in Fig.\ref{fig:nd-all}b and a superlattice reflection at $(0,0,1.5)$ is observed throughout the antiferromagnetic phase (cf. Fig.\ref{fig:nd-all}c). Note that in contrast to the other compounds the crystal symmetry remains $R32$ down to lowest temperature.

In contrast to these previous cases, in \tb~ the Fe moments are parallel to the $c$-axis and no superlattice reflection is present in the whole zero magnetic field phase at $T<T_N=39$~K. Thus, the afore discussed conjecture of in-plane oriented Fe-moments as the required condition for observing the superlattice peaks is further substantiated.

\subsection{Field dependence of the (0,0,1.5) reflection}
\begin{figure}
 \includegraphics[angle=-90,width=\columnwidth]{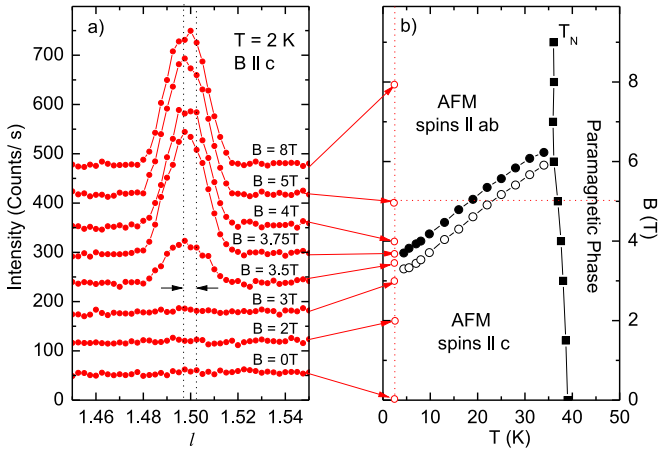}
 \caption{Hard x-rays diffraction measurements on \tb. Plot a) shows the evolution of the (0,0,1.5) reflection as the applied magnetic field is increased from 0 to 8T. Dashed lines refer to the FWHM of the (0,0,3) reflection, which was used as the experimental resolution function in order to estimate the correlation length of the magnetic signal (read text). b) shows the phase diagram reproduced from Ref.~\onlinecite{Popova07a}.}
 \label{fig:PHD}
\end{figure}

\begin{figure}
 \includegraphics[width=\columnwidth]{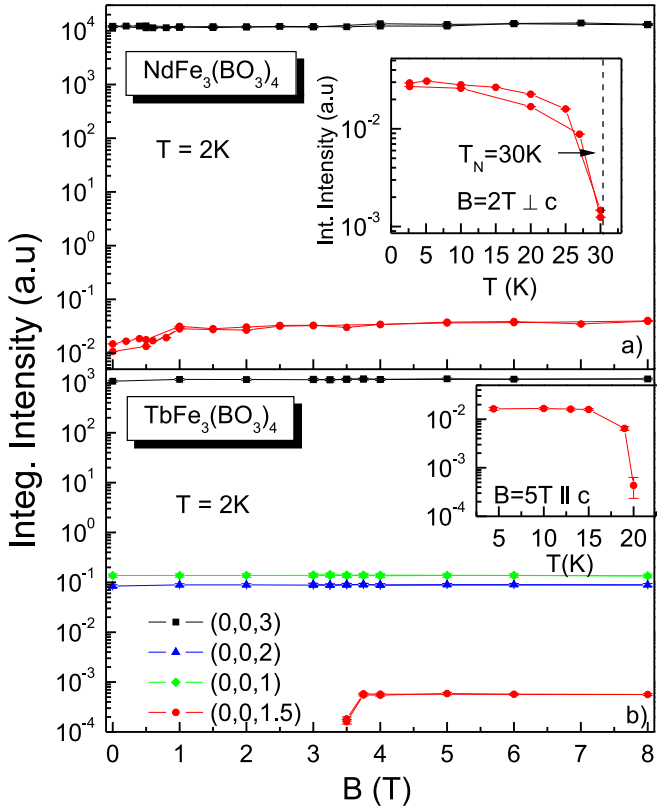}
 \caption{Hard x-rays diffraction integrated intensities as a function of magnetic field for the reflections (0,0,1), (0,0,1.5), (0,0,2) and (0,0,3) measured on a) \nd~ and b) \tb. The insets show the evolution of the magnetic (0,0,1.5) peak at a given applied magnetic field, as a function of temperature.}
 \label{fig:nd-tb-field}
\end{figure}

In order to manipulate the Fe spin orientation and thereby investigate the field dependence of the $(0,0,1.5)$ reflection we applied external magnetic fields. In \tb, which exhibits AFM order with $c$ as the easy axis in zero field, the metamagnetic transition which occurs upon the application of an external magnetic field parallel $c$ induces an in-plane configuration of the iron spins\cite{Popova07a}. Fig.~\ref{fig:PHD}b reproduces the magnetic phase diagram of this compound from Ref.~\onlinecite{Popova07a} which allows to elucidate the temperature and magnetic field dependence of the metamagnetic transition. Fig.~\ref{fig:PHD}a shows that at low temperature $T=2$~K the $(0,0,1.5)$ peak emerges from the background at magnetic fields $B\approx3.5$~T, i.e., as soon as the Fe moments are oriented in-plane, it quickly gains intensity up to saturation at higher magnetic fields (see Fig.~\ref{fig:nd-tb-field}b). The peak width does not change when increasing the field, supporting the picture of a spin-flop where there is no significant change of the spin correlation length, but only of the spin direction. The inset in figure \ref{fig:nd-tb-field}b shows the intensity of this peak as a function of temperature at a fixed external field of $B=5$~T parallel to $c$, which corresponds to a horizontal cut through the phase diagram in Fig. \ref{fig:PHD}. As can be seen in the figure, the peak intensity is constant at $T\leq15$~K, rapidly decreases at higher $T$ and eventually vanishes at $T\gtrsim 20$~K. Note, that upon crossing this temperature the Fe-moments reorient from the in-plane to the parallel $c$ configuration of the AFM phase. 
We have performed a similar measurement also for \gd. As has been discussed already above, in zero magnetic field, the $(0,0,1.5)$ reflection appears only in the temperature range between $T_\mathrm{SR}$ and $T_N$ (Figure \ref{fig:gd-all}b). According to Yen et al.\cite{yen06} the spin reorientation temperature $T_\mathrm{SR}$ decreases if a magnetic field parallel to the $c$ direction is applied. As can be seen in Fig.~\ref{fig:gd-all}b for the case of $B=0.6$~T, the downshift of $T_\mathrm{SR}$ also leads to a shift of the lowest temperature at which the $(0,0,1.5)$ reflection appears. To be specific, at the applied magnetic field the reflection is well resolved at $T\geq7$~K, i.e. in perfect agreement with the phase diagram reported by Yen et al~\cite{yen06}.

\begin{figure}
 \includegraphics[angle=-90,width=\columnwidth]{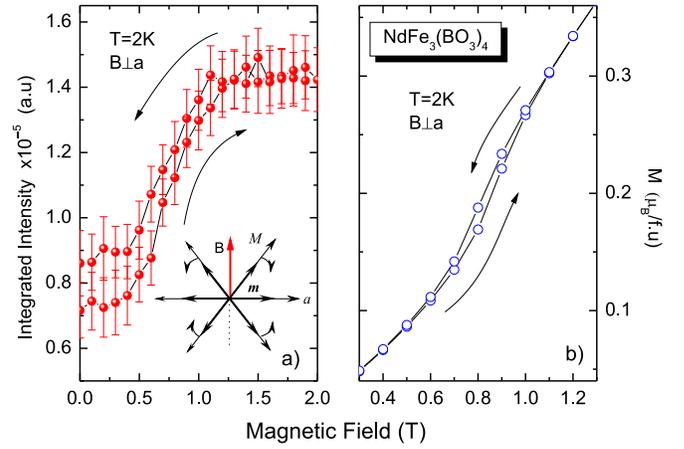}
 \caption{a)Integrated intensity of the magnetic (0,0,1.5) reflection as a function of magnetic field. The inset shows the different spin orientations along the equivalent directions ($a_1, a_2, a_3$) in the basal plane and how they turn when a magnetic field is applied along a direction perpendicular to $a_1$ in the basal plane\cite{volkov07-2}. $M$ and $m$ corresponds to the Fe and Nd moments respectively. b) Magnetization measurement of \nd~ at low values of magnetic field. The magnetic field was applied perpendicular to the crystallographic $a$ direction along the basal plane. Note that the magnetic peak mimics the hysteretic behavior shown in magnetization.}
 \label{fig:spin-flop}
\end{figure}

Both examples where the $(0,0,1.5)$ superlattice reflection is induced by an external magnetic field unambiguously demonstrate that there is  a clear one-to-one correlation between this reflection and the in-plane orientation of the Fe spins. In the following we point out that even more subtle changes of the magnetic structure have an impact on the intensity of the reflection.
Figure \ref{fig:nd-tb-field}a shows the integrated intensity of the magnetic reflection as a function of magnetic field for \nd~ with the magnetic field oriented in-plane. It can be seen that the magnetic peak intensity increases with increasing the field, until it reaches saturation at $B\sim1$ T. Results of a detailed measurement in this field range are shown in figure \ref{fig:spin-flop}a and reveal that the peak intensity exhibits a steep increase at $0.5~{\rm T}\lesssim B\lesssim 1.2$~T, while it is constant at lower and higher fields. Interestingly this increase of peak intensity coincides with an increase of magnetization as is shown in Fig.~\ref{fig:spin-flop}b. The small jump in magnetization is the signature of a spin flop of both Fe and Nd moments, along a direction perpendicular to the external field within the basal plane\cite{volkov07-2}.

\subsection{Discussion}
\label{sec:discussion}
The magnetic structure of \re~ ($\rm R= Y$, Nd, Tb) has been studied in a number of different neutron diffraction experiments\cite{fischer06,ritter07,ritter08}. Depending on the compound, either a spin spiral (\nd) propagating along the $c-$axis\cite{fischer06} or a N\'{e}el state\cite{ritter07,ritter08} (\yfe~ and \tb) have been inferred from the data. In all cases commensurate magnetic superlattice reflections occur at $(0,0,1.5)$, i.e. the magnetic supercell is doubled along the $c-$axis. Hence, at first glance two possible origins of the x-ray superlattice reflections should be considered. Firstly, the observed superlattice reflections could be of \textit{magnetic} nature and thus represent the x-ray pendant of the magnetic reflections seen in neutron scattering. Secondly, the superlattice reflections could result from a weak structural distortion that is imposed by the magnetic ordering. An appealing scenario for the latter which reasonably explains the observed correlation between the occurrence of the superlattice reflections and the orientation of the Fe-spins is a magnetic spiral state which propagates along the $c$-axis (as is concluded from neutron data for \nd\cite{fischer06}) and thus could in principle lead to a structural distortion following the spin spiral. This spiral state (including the assumed lattice distortion) requires the Fe moments or a component of them to lie in the $ab$-plane and thus it has to vanish as soon as the Fe moments are fully oriented along $c$.\\
On the other hand, there are compelling reasons to rationalize the $(0,0,1.5)$ superlattice reflection in terms of purely magnetic scattering. The Bragg angle for this reflection at $h\nu=100$~keV is $\theta_{(0,0,1.5)}=0.705^\circ$, i.e. close to $\theta\rightarrow 0$. As discussed in section \ref{sec:cross-sect}, the cross section for magnetic scattering under this experimental conditions allows to observe magnetic scattering when the spins of the unpaired electrons have a component perpendicular to the scattering plane (eq. \ref{eq:magnetic-matrix-simple}). In our experimental setup this is indeed the case if the iron spins lie in the $ab$ plane of the sample. Thus, the surprising dependence of the peak intensity on the spin orientation is naturally explained without further assumptions. Moreover, the intensity of this reflection is around seven orders of magnitude weaker than the $(0,0,3)$ Bragg peak, which is just the expected order of magnitude for magnetic reflections (cf. equation~\ref{eq:magn-charg-ratio}). Thus it seems more reasonable to assign a magnetic origin to the superlattice reflection.
\subsection*{q and azimuth dependence of the magnetic reflection}
\begin{figure}
\includegraphics[angle=-90,width=\columnwidth]{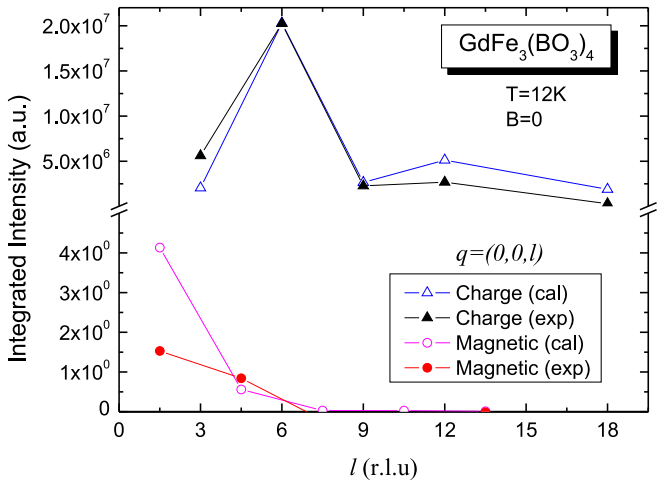}
\caption{Intensity of the diffracted peaks as a function of \textbf{q} for the \gd~ as measured (exp, filled symbols) and calculated (cal, open symbols). The calculated intensities from the structure factors were normalized with respect to the observed reflections for comparison. The normalization factor is the same for both the structural and the superlattice reflections.}
\label{fig:q-scans}
\end{figure}
In order to verify the latter conclusion we have performed measurements of the  $\textbf{q}$-dependence of the intensity of both the structural $(0,0,3l)$ Bragg reflections and of the $(0,0,3l\pm1.5)$ reflections, where we studied \gd~ as a representative case (see Fig.~\ref{fig:q-scans}). As a function of $\textbf{q}$, the measured intensity of the structural Bragg reflections (filled triangles) evolves as it is expected from structure factor calculations (open triangles). One can see that even at large $\textbf{q}$ values the charge reflections are still strong. 
A much faster decrease of intensity with increasing $\textbf{q}$ is expected for magnetic reflections as is shown in the figure by the open diamonds, which agrees well with the observed intensities of the $(0,0,3l\pm1.5)$ reflections and thus allows to unambiguously identify the peak as magnetic.
\footnote{The charge and magnetic form factors were calculated using the tabulated magnetic and charge form factors reported in the international tables of crystallography\cite{brown}. The atom positions were obtained from single crystal x-ray diffraction (SCXRD) measurements performed at T=100K, using a Bruker Kappa APEX II diffractometer with Mo $K_{\alpha1}$ radiation. The data were refined using the program SHELX\cite{sheldrick}. The magnetic structure factor was calculated taking into account only the Fe atoms, since these are the ones which have the major contribution to the magnetization, and assuming $S=5/2$.} 

\begin{figure}
\includegraphics[angle=-90,width=\columnwidth]{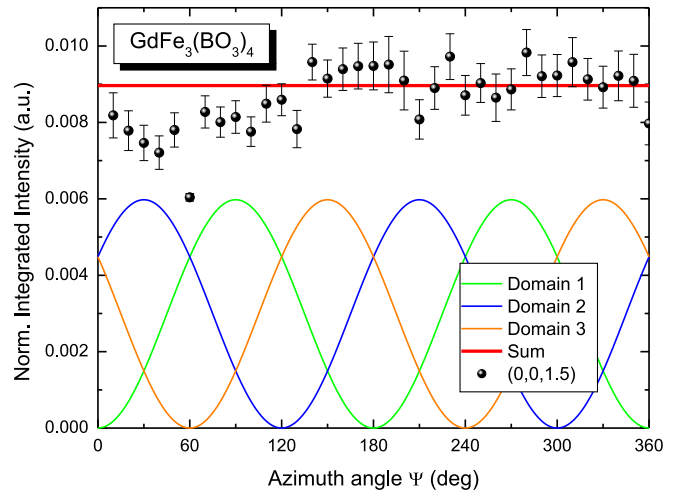}
\caption{Calculated (lines) and experimentally observed (points) azimuth dependence of the magnetic (0,0,1.5) reflection in \gd. The magnetic intensities were normalized with respect to the (0,0,3) Bragg reflection.}
\label{fig:gd-azimuth}
\end{figure}

In principle, the peak intensity at $(0,0,1.5)$ should depend on the azimuth angle if an easy axis exists when the Fe spins lie within the $ab$ plane. Measurements at different azimuth angles at $\textbf{q}=(0,0,1.5)$ are presented in Fig.~\ref{fig:gd-azimuth}. Since the integrated intensity of the magnetic peak shows a constant azimuth dependence, only two possible scenarios are in agreement with the experiment. The first possible interpretation involves an equally populated domain structure, where the three equivalent domains in the basal plane are rotated 120$^\circ$ relative to each other. Figure \ref{fig:gd-azimuth} shows the expected signal for the three different domains and the red line refers to the summation of the three domain signals. The second possible reason of such azimuth dependence is the formation of a spin spiral as suggested from neutron diffraction data\cite{fischer06}. 

It is interesting to point out in this regard the afore described effect of an external magnetic field on the $(0,0,1.5)$ reflection where the field was applied in-plane. Since in our setup the field direction lies in the scattering plane, a spin flop of the iron moments will definitely manipulate the number of spins perpendicular to the scattering plane. More specifically in \nd, following a spin model suggested by Volkov et al,\cite{volkov07-2} one should expect three equivalent easy directions of magnetization in the basal plane (see inset in fig. \ref{fig:spin-flop}a). The applied magnetic field causes the spins to flop into a direction perpendicular to the applied field, i.e perpendicular to the x-ray scattering plane. From equation \ref{eq:magnetic-matrix-simple} it is clear that an increase of the spin component perpendicular to the scattering plane enlarges the magnetic scattering cross section and therefore yields an increase of the intensity. Figure \ref{fig:spin-flop}a actually shows the enhancement of the magnetic intensity due to the rotation of the magnetic moments along the scattering plane normal. 
\subsection*{Correlation length $\xi$ of the magnetic domains}
The width of the magnetic reflection is around twice the width of the nearest Bragg reflection, as shown by the dashed lines in figure \ref{fig:PHD}a for \tb. This suggests that the correlation length, or size of the magnetic domains within the crystal, has a finite size. Experimentally, the measured diffracted signal is a convolution of the experimental resolution function and the intrinsic diffracted signal from the crystal. Hence, the measured signal $h$ can be expressed as:
\begin{equation}
[f \ast g](t)=h(\tau)
\label{eq:convolution}
\end{equation}
Where $[f \ast g]$ denotes the convolution of the experimental resolution function $f$ and the intrinsic signal $g$.
If we consider the Gaussian fit profile of the (0,0,3) Bragg reflection as our experimental resolution function where FWHM$_{(0,0,3)}$ = 0.0072(1) r.l.u., an estimation of the intrinsic magnetic diffracted signal can be obtained by a deconvolution from the measured signal. As the measured signal can also be expressed by a Gaussian, thus the intrinsic function $g$ is also a Gaussian with full width at half maximum (FWHM) equal to:
\begin{equation}
\Delta_g=\sqrt{\Delta_h^2-\Delta_f^2} 
\label{eq:FWHM}
\end{equation}
After deconvolution, the correlation length of the magnetic domains along the $c$ direction was estimated assuming a perfect crystal with no strain, assumption supported by the small crystal mosaicity, 0,0060(1)$^{\circ}$, determined from the 2$\theta$ scan of the (0,0,3) Bragg reflection. For all compounds we find $\xi_c\approx100\AA{}$ (cf. table \ref{tab:table1} for details). The correlation length $\xi_c$ was calculated using the following relation: 
\begin{equation}
\frac{1}{\xi_c}=2\pi\left|\left(0,0,\frac{\Delta_g}{c}\right)\right|
\label{eq:correlation-length}
\end{equation}
\begin{table}[h]
\caption{Correlation length of the magnetic domains, obtained from the FWHM of the deconvoluted magnetic signal.}
\begin{ruledtabular}
\begin{tabular}{lcc}
Sample&$\xi_c$(\AA{})& \\
\hline
\yfe & 107 $\pm$ 33 & T=5 K \\
\gd& 99 $\pm$ 5 &  T=10 K  \\
\tb& 101 $\pm$ 1 & T=2 K, B $>$ 3.5 T \\
\nd& 93 $\pm$ 2 &  T=2 K, B=0  \\
\nd& 104 $\pm$ 2 & T=2 K, B $>$ 2 T 
\label{tab:table1}
\end{tabular}
\end{ruledtabular}
\end{table}
\subsection*{The magnetic structure factor and the spin-moment $S$ in GdFe$_3$(BO$_3$)$_4$}

Finally, according to equation \ref{eq:mag-cross-section} it is possible to determine the size of the component of $S$ perpendicular to the scattering plane from the measured integrated intensity of a magnetic reflection.
In the kinematical approximation, the magnetic and charge reflectivity ratio  of a crystal in a Laue symmetry has the following form:
\begin{equation}
\frac{R_m}{R_c}=\frac{I_m}{I_c}=\left(\frac{\lambda_{c}}{d}\right)^{2}\frac{\sin\theta^{c}_{B}}{\sin\theta^{m}_{B}}\left(\frac{|F_m|}{|F_c|}\right)^{2}k
\label{eq:reflectivity}
\end{equation}
$I_m$, $I_c$, $|F_m|\varpropto S_{2}$ and $|F_c|$ are the measured integrated intensities of the magnetic and charge reflections, and the magnetic and  charge structure factors, respectively. $\theta^{m}_{B}$ and $\theta^{c}_{B}$ are the Bragg angles of the magnetic and charge scattering, $d$ is the interatomic spacing of the magnetic reflections and $k=3$ is a correction factor that compensates for the three equally populated domains in which the magnetic moment lies in the hexagonal basal plane. 
The direct measurement of the intensity of the $(0,0,3)$ Bragg peak is not possible since the strong intensity saturates the detector. It is necessary to measure this peak at different radiation absorbers and finally extrapolate to absorber 0.
Using the calculated value of $|F_c|$ and knowing that only the iron atoms contribute to the magnetic scattering in the sample, the solution of equation \ref{eq:reflectivity} for $|F_m|$ yields an estimated value of $S_{2}=1.49\pm0.09$. Since this values is a projection of $S$ along the scattering plane normal, which coincides with the $ab$ plane as discussed before, one can calculate the angle between the iron moments and the basal plane. Taking $S=5/2$ as reported from magnetization measurements\cite{yen06}, this angle is $53^\circ\pm3^\circ$, which is in close agreement with the angle found by resonant scattering experiments\cite{mo} which is around $45^\circ$. Deviation from this values could arise from experimental restrictions, since the determination of the primary intensity of the (0,0,3) reflection is indirect and also that the weakness of the magnetic intensity can bring some systematic error while data acquisition.

\section{Summary}
\label{sec:summary}
Structural and magnetic properties on compounds of the form \re~ ($\rm R = Gd$, Tb, Nd and Y) have been studied by means of high energy x-ray diffraction. Due to the high energy photons used during the experiment, NRXMS could be observed on all the compounds at temperatures below the magnetic order temperature when the AFM vector lies in the $ab$ plane. The study of the magnetic reflection as a function of temperature and applied magnetic field shows the different metamagnetic transitions such as spin flops and spin reorientation transitions in \tb, \nd~ and \gd. Moreover NRXMS allowed us to corroborate the magnetic structures obtained from neutron scattering experiments. Detailed analysis of the magnetically diffracted intensities  as a function of magnetic field and azimuth angle in \nd~ and \gd~ respectively, suggests that the magnetic moments of the Fe ions are aligned in the crystallographic $a$ axis, leading to a domain structure formation since there are three equivalent directions in the hexagonal basal plane. For \gd~we extracted an out-of-plane angle of $53^\circ\pm3^\circ$ for the iron moments and for all the compounds, a correlation length of the magnetic domains of $\sim$100 \AA{} was estimated. 
 
Furthermore, we observed the presence of new superlattice reflections at $(0,0,3l\pm1)$. These suggest that the symmetry of the crystal has not been properly assigned, as these reflections violate the reflection conditions for the until now accepted $P3_121$ space group.

\section{Acknowledgments}

This work was supported by the Deustche Forschungsgemeinschaft, through the Forschergruppe FOR520 (grant HE3439/6) and HASYLAB at the Deutsches Elektronen-Synchrotron (DESY). 

The authors would like to acknowledge Dr. Sergio Valencia and Dr. Ralf Feyerherm from HMI Berlin and specially to Dr. J\"{o}rg Strempfer from HASYLAB at DESY in Hamburg for valuable discussions.

\end{document}